\documentclass[10pt]{article}
\usepackage{graphicx}
\usepackage{amsmath}
\usepackage{lineno}

\def\Title#1{\begin{center} {\Large #1 } \end{center}}
\def\Author#1{\begin{center}{ \sc #1} \end{center}}
\def\Address#1{\begin{center}{ \it #1} \end{center}}

\newcommand\pubblock{\rightline{\begin{tabular}{l} Proceedings of the Fifth Annual LHCP\\ \pubnumber\\
         \pubdate  \end{tabular}}}

\newenvironment{Abstract}{\begin{quotation} \begin{center} 
             \large ABSTRACT \end{center}\bigskip 
      \begin{center}\begin{large}}{\end{large}\end{center} \end{quotation}}

\newenvironment{Presented}{\begin{quotation} \begin{center} 
             PRESENTED AT\end{center}\bigskip 
      \begin{center}\begin{large}}{\end{large}\end{center} \end{quotation}}

\def\Acknowledgements{\bigskip  \bigskip \begin{center} \begin{large}
             \bf ACKNOWLEDGEMENTS \end{large}\end{center}}

\def\beq{\begin{equation}}
\def\eeq#1{\label{#1}\end{equation}}
\def\eeqn{\end{equation}}

\def\beqa{\begin{eqnarray}}
\def\eeqa#1{\label{#1}\end{eqnarray}}
\def\eeqan{\end{eqnarray}}

\let\bar=\overbar

\def\Dslash{\not{\hbox{\kern-4pt $D$}}}
\def\dslash{\not{\hbox{\kern-2pt $\del$}}}

\def\BR{\mathcal{B}}
\def\pt{p_\mathrm{T}}
\def\gevc{\mathrm{GeV}\!/c}

\def\Lcp{\Lambda_c^+}
\def\Lb{\Lambda_b^0}
\def\Bp{B^+}
\def\Bd{B^0}
\def\Bs{B_s^0}
\def\jpsi{J\!/\psi}
\def\FLbFd{f_{\Lb}/f_d}

\def\msb{{\bar{\ssstyle M \kern -1pt S}}}

\usepackage{color}

\newcommand\pubnumber{ LHCb-PROC-2017-035} 

\newcommand\pubdate{\today}

\def\affiliation{
On behalf of the LHCb Collaboration, \\
LAL, Universit\'e Paris-Sud, CNRS/IN2P3
Orsay, 91898, France}

\begin{document}

\large
\begin{titlepage}
\pubblock

\vfill
\Title{Measurement of heavy flavor production in $pp$ collisions at LHCb}
\vfill

\Author{ Yanxi ZHANG}
\Address{\affiliation}
\vfill
\begin{Abstract}
Heavy flavor productions are important tests of QCD. In proton-proton collisions collected at LHCb, a long list of measurements for charm and bottom
productions have been made. This talk focuses on the studies of fragmentation functions and production asymmetries for
bottom hadrons. They provide inputs for absolute branching fraction measurements and CP violation studies at LHCb respectively.

\end{Abstract}
\vfill

\begin{Presented}
The Fifth Annual Conference\\
 on Large Hadron Collider Physics \\
Shanghai Jiao Tong University, Shanghai, China\\ 
May 15-20, 2017
\end{Presented}
\vfill
\end{titlepage}
\def\thefootnote{\fnsymbol{footnote}}
\setcounter{footnote}{0}
%

\normalsize 

\section{Introduction}
The hadronic production of heavy flavor hadrons involves heavy quark production from partonic interactions and
then the hadronization, both described in the framework of QCD. A wide range of measurements has been performed by
the LHCb experiment including the charm and bottom hadron production cross-sections, the quarkonium productions and
polarizations, associated productions and production correlations. The measurements are important to understand the QCD theory
in both the perturbative and non perturbative parts. Currently QCD is not fully understood, especially the
non-perturbative part: there is no universal model that could describe all heavy flavor productions. 
In the talk, I summarized the LHCb measurements on $b$-hadron
fragmentation functions and production asymmetries. 
Fragmentation functions stand for the relative production rates of different $b$-hadrons, $f_q$ for the $B_q$ hadron. 
The production asymmetries measure the difference of production rates between a $b$-hadron and its charge conjugate.
These processes involve non-perturbative effects, and thus 
reply on experimental data to fix model parameters. The fragmentation fractions determine the total production yield for each
$b$ hadron species, providing inputs to calculate absolute branching fractions. The $b$ hadron production asymmetries
are essential inputs for measurements of CP asymmetries at LHCb.

\section{Measurements of $b$-hadron fragmentation functions}
\begin{figure}[htb]
\centering
\includegraphics[width=0.49\textwidth]{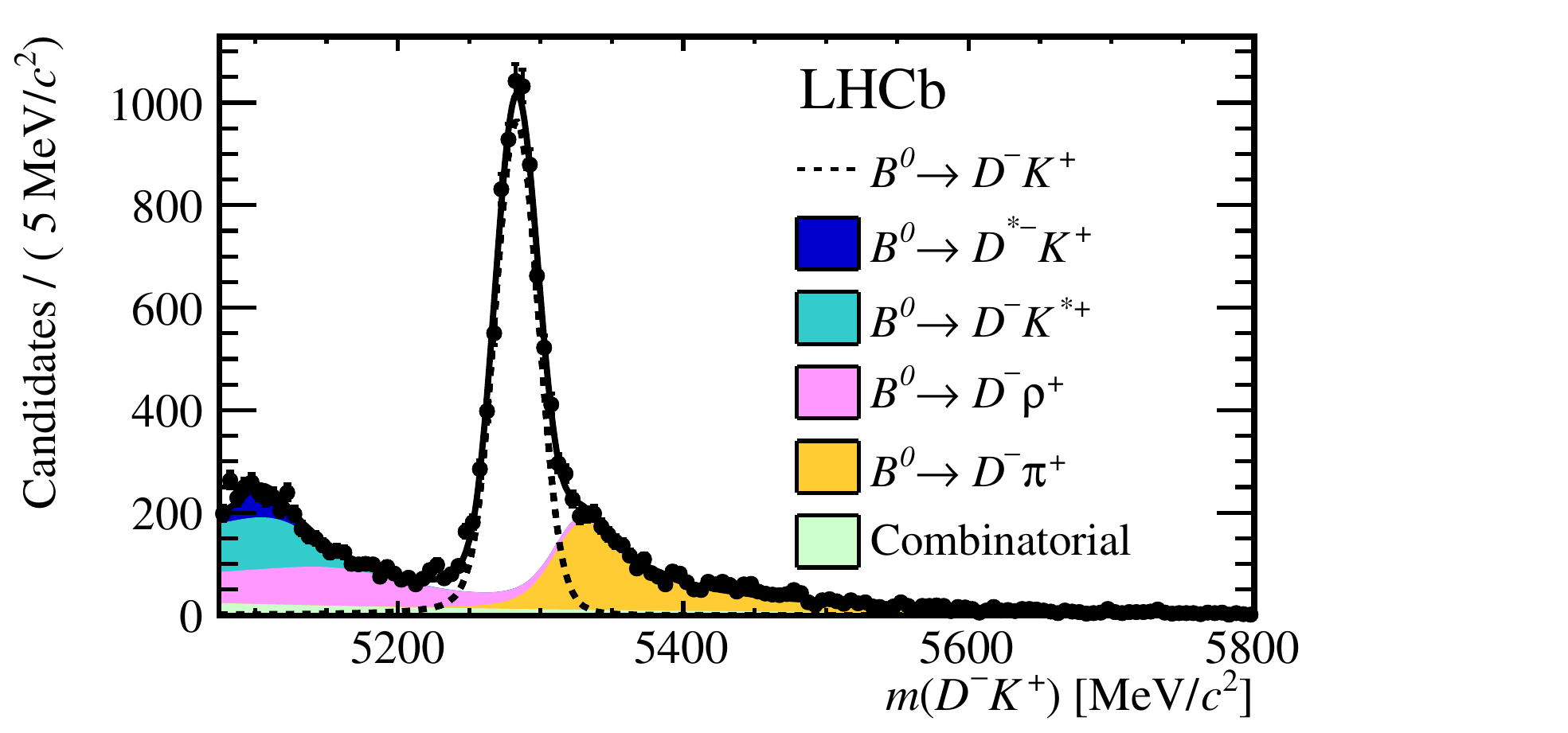}
\includegraphics[width=0.49\textwidth]{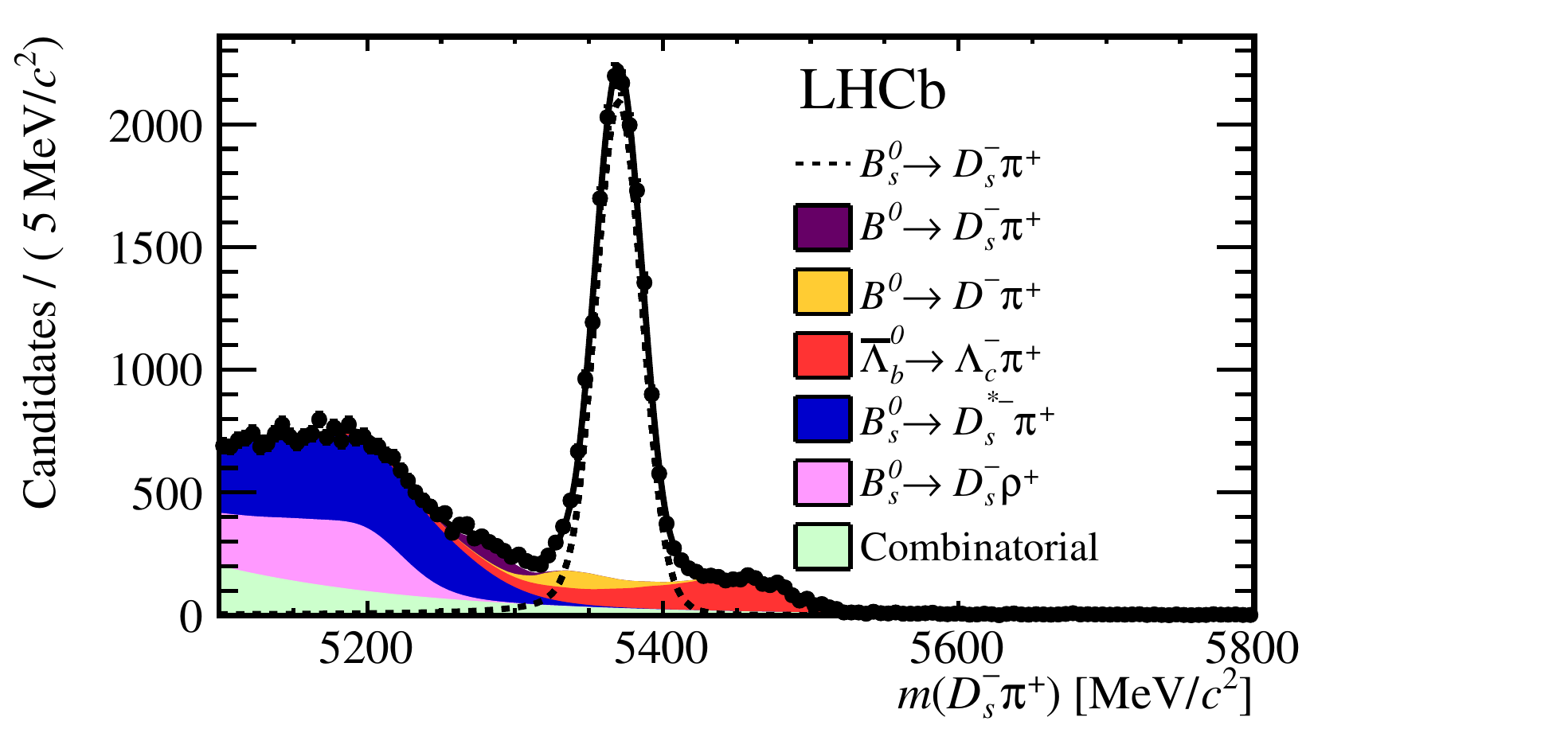}
\caption{ Invariant mass distributions of (left) $m(D^-K^+)$ and (right) $m(D_s^-\pi^+)$. 
}
\label{fig:figure1}
\end{figure}

\begin{figure}[htb]
\centering
\includegraphics[width=.40\textwidth]{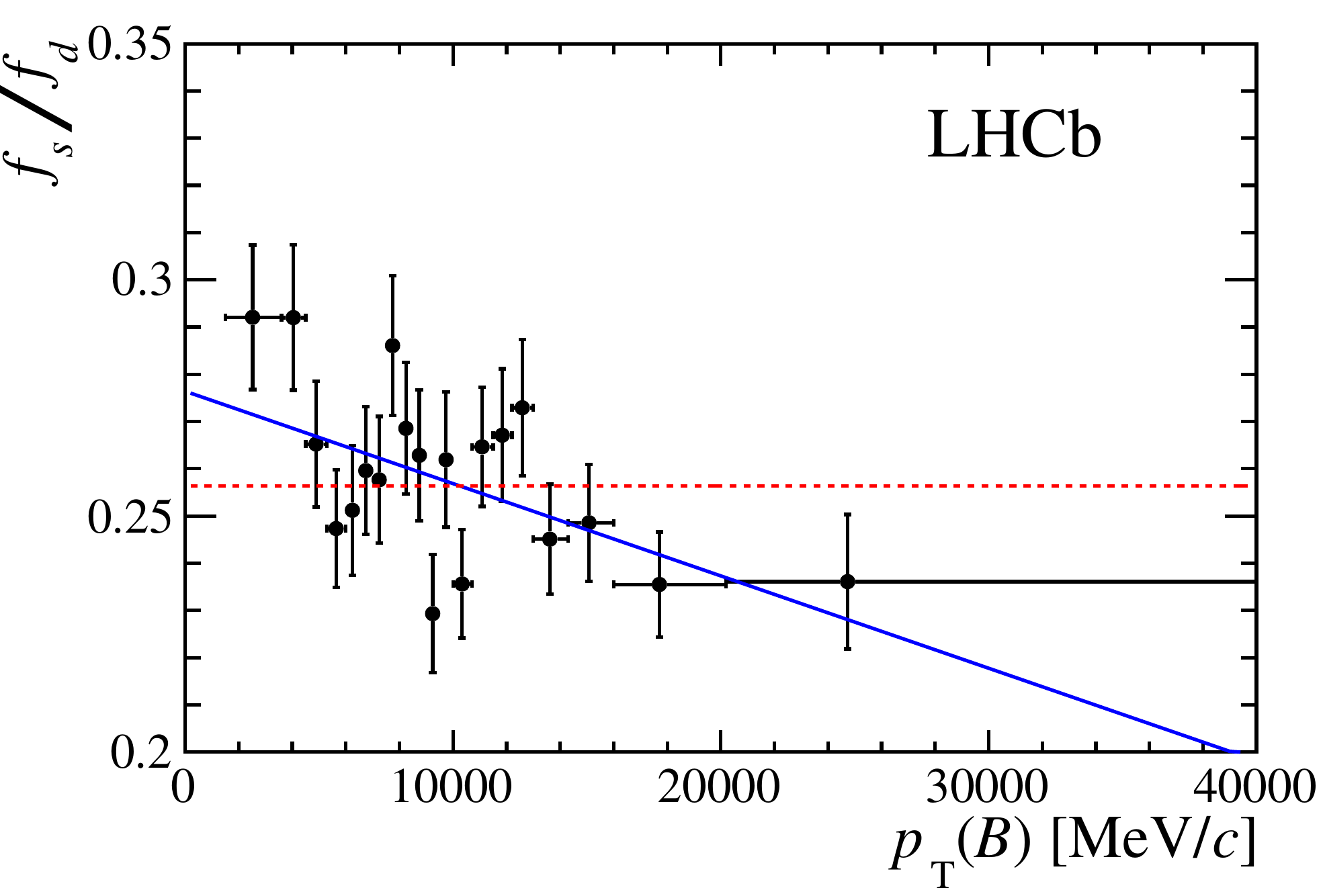}
\includegraphics[width=.40\textwidth]{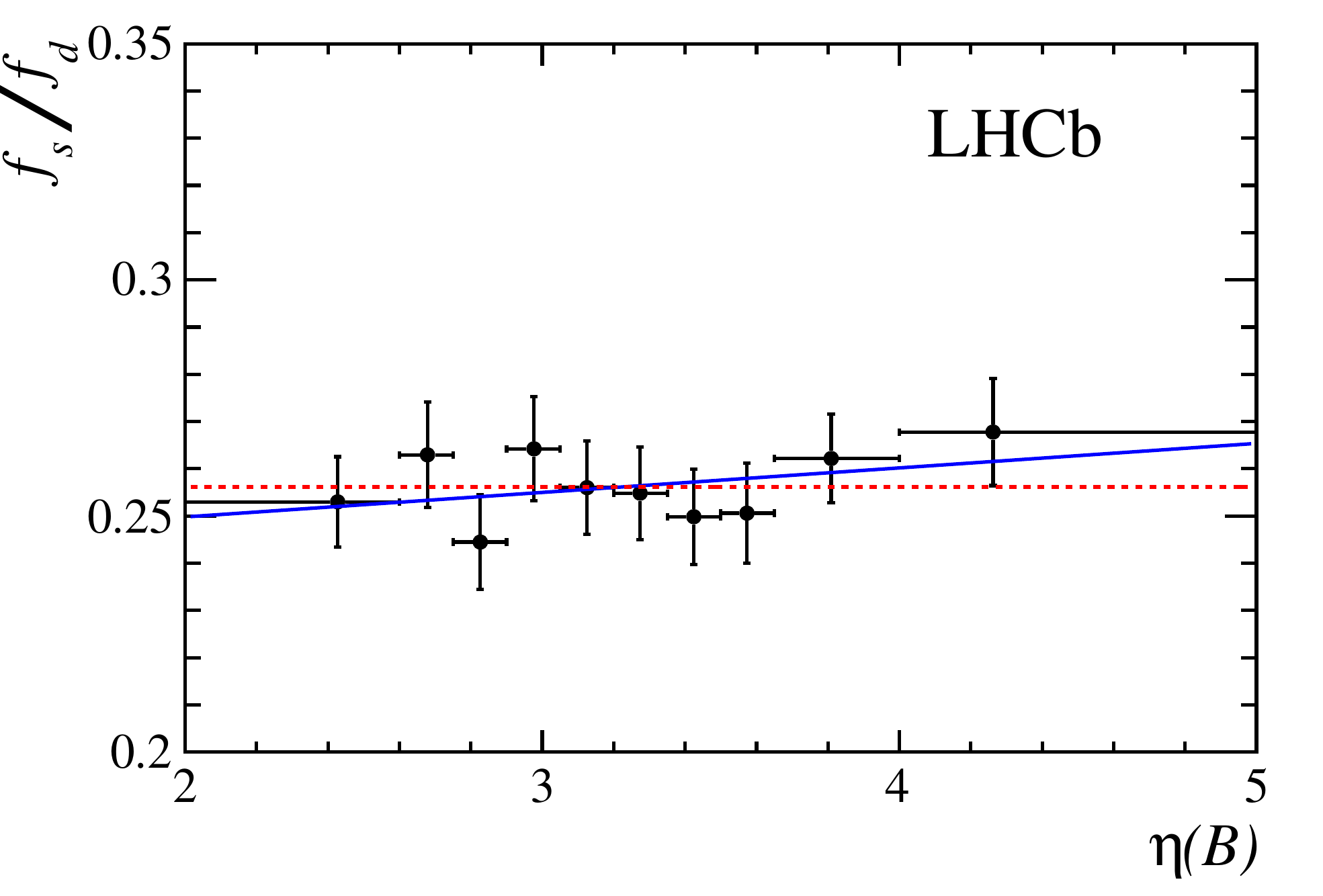}
\caption{ Ratio of fragmentation fractions $f_s/f_d$ as functions of (left) $\pt$ and (right) $\eta$ of the $b$-hadrons.
The solid blue line represents the linear fit, while the dashed red line corresponds to the average.}
\label{fig:figure2}
\end{figure}

LHCb measured the relative fragmentations between $B_s^0$ and $B_0$, $f_s/f_d$, using the decays $B_s^0\to D_s^-\pi^+$ and
$B^0\to D^-K^+$~\cite{cite1} in $pp$ collisions at $\sqrt{s}=7$ TeV. Experimentally $f_s/f_d$ is determined as $\frac{f_s}{f_d}=\frac{\BR(B^0\to D^-K^+)}{\BR(B_s^0\to
D_s^-\pi^+)}\frac{\epsilon_{D^-K^+}}{\epsilon_{D_s^-\pi^+}}\frac{N_{D_s^-\pi^+}}{N_{D^-K^+}}$. 
The two decay modes 
both receive only contributions of color-allowed tree-diagram and are connected by U-spin symmetry. The branching fraction
ration $\frac{\BR(B^0\to D^-K^+)}{\BR(B_s^0\to D_s^-\pi^+)}$ was calculated using phenomenological model assuming
factorization, including small non-factorizable U-spin-breaking corrections. The $D^-$ and $D_s^-$ mesons are
reconstructed in the $D^-\to K^+\pi^-\pi^-$ and $D_s^-\to K^+K^-\pi^-$ decays respectively. The same event selection criteria were
applied to the two channels, including the trigger requirement that selects a high transverse momentum track displaced
from the primary vertices, the particle identification requirements on the final state tracks and a multivariate
selector trained using kinematic and geometrical variables. The signal yields $N_{D_s^-\pi^+}$ and $N_{D^-K^+}$ are
obtained from the fit to the invariant mass distributions, as shown in Figure~\ref{fig:figure1}. 
The two decay modes have similar decay topology, largely cancelling the systematic uncertainties on the efficiencies. 
The systematic uncertainty on the $f_s/f_d$ measurement consists of that on signal yields and those on trigger and offline selection
efficiencies, giving a total value of 3.4\%. Another uncertainty comes from the branching fraction ratio of $D_s^+$ and
$D^-$ decays. Averaged over LHCb acceptance,
$f_s/f_d$ is measured to be $0.238\pm0.004\pm0.015\pm0.021$,
where the uncertainties are statistical, systematic uncertainties excluding that of $\frac{\BR(B^0\to
D^-K^+)}{\BR(B_s^0\to D_s^-\pi^+)}$ and that of $\frac{\BR(B^0\to D^-K^+)}{\BR(B_s^0\to D_s^-\pi^+)}$. 
The $B^0\to D^-\pi^+$ channel, which has ten times more yields compared to $B_0\to D^-K^+$, is used to measure kinematic
dependent $f_s/f_d$, for which the result is displayed in Figure~\ref{fig:figure2} as functions of the $b$-hadron $\pt$ and $\eta$.
Fits using linear functions are performed giving
\begin{eqnarray}
    f_s/f_d (\pt)&=& (0.256\pm0.020) + (-2.0\pm0.6) \times 10^{-3}/(\gevc)\times (\pt-10.4\,\gevc)\nonumber\\
    f_s/f_d (\eta)&=& (0.256\pm0.020) + (0.005\pm0.006) \times (\eta-3.28). \nonumber
\end{eqnarray}
The $f_s/f_d$ slightly  depends on $\pt$ with a significance of three standard deviations, while no
indication of a dependence on $\eta(B)$ is found. The measurement concludes that the production rate of $B_s^0$ meson is
approximately $25\%$ of that of $B^0$ at LHCb. The result provides the inputs for branching fraction measurements for
$B_s^0$ using $B^0$ decays as references. 

\begin{figure}[htb]
\centering
\includegraphics[width=.38\textwidth]{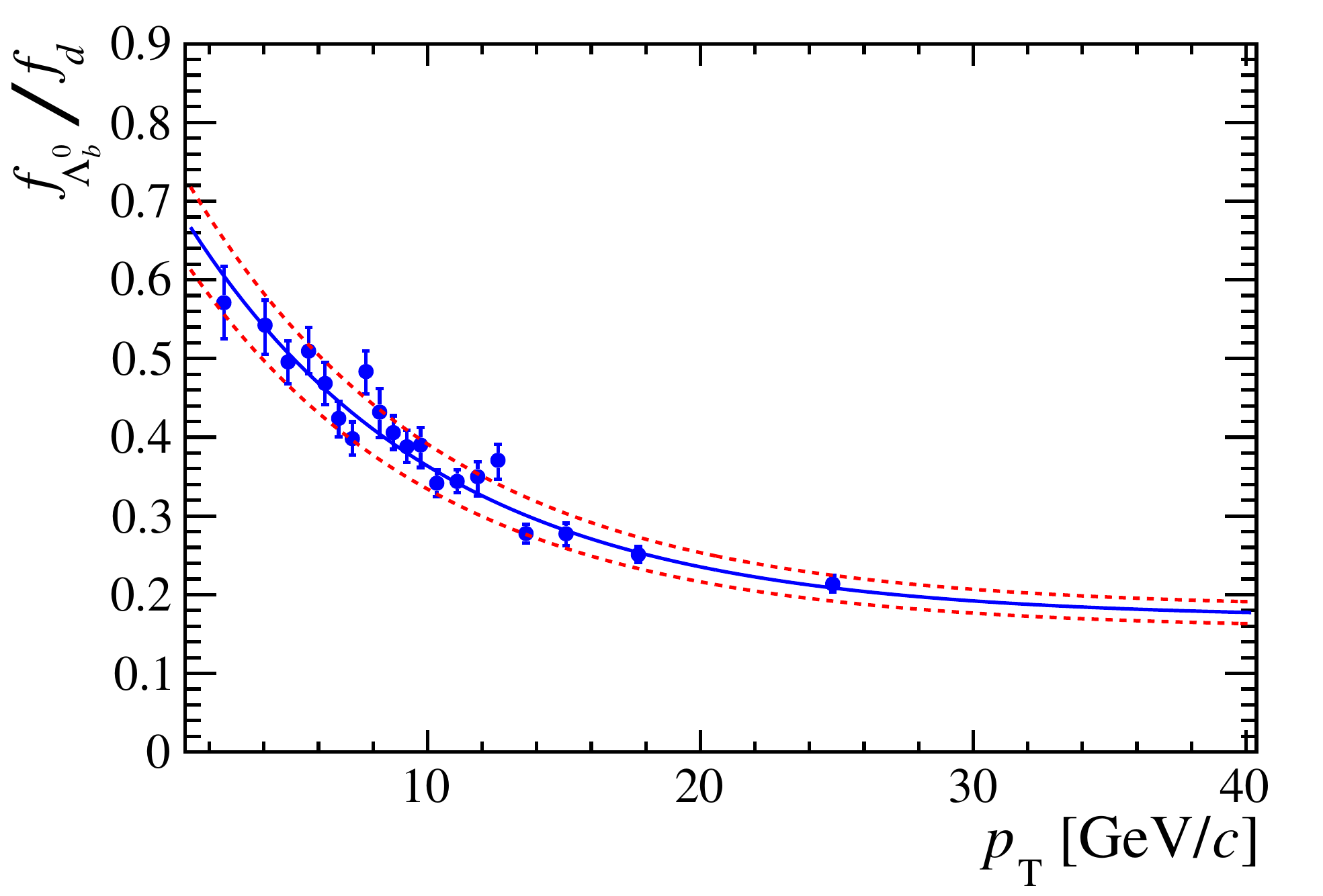}
\includegraphics[width=.38\textwidth]{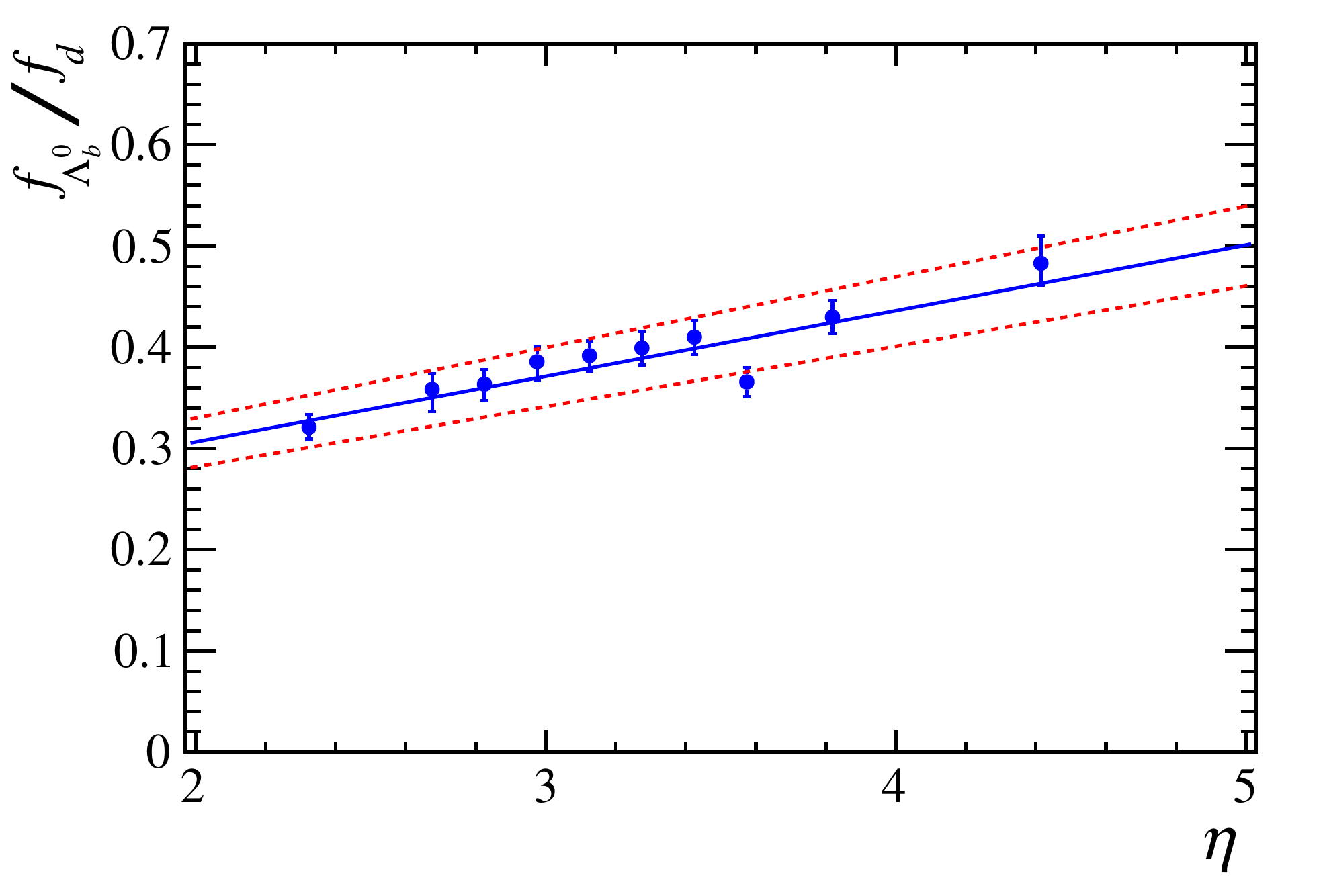}
\caption{ 
    Dependence of $\FLbFd$ on the (left) $\pt$ and (right) $\eta$ of the beauty hadron. 
    The solid line represents the fit with exponential and linear function respectively, while the dashed lines
    corresponds to one standard deviations on the scale of $\FLbFd$ extrapolated from semileptonic analysis.
}
\label{fig:figure4}
\end{figure}

\begin{figure}[htb]
\centering
\includegraphics[width=.40\textwidth]{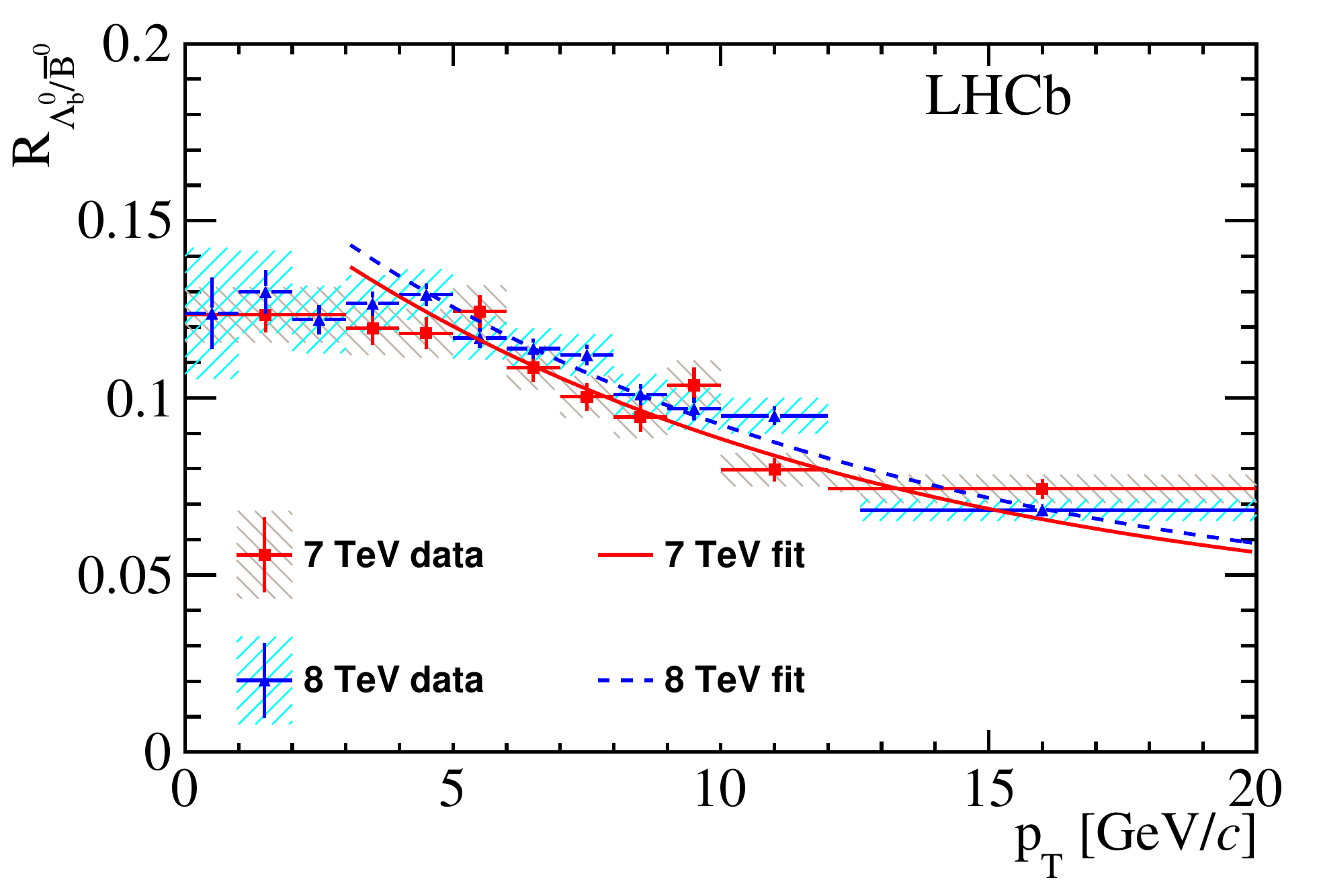}
\includegraphics[width=.40\textwidth]{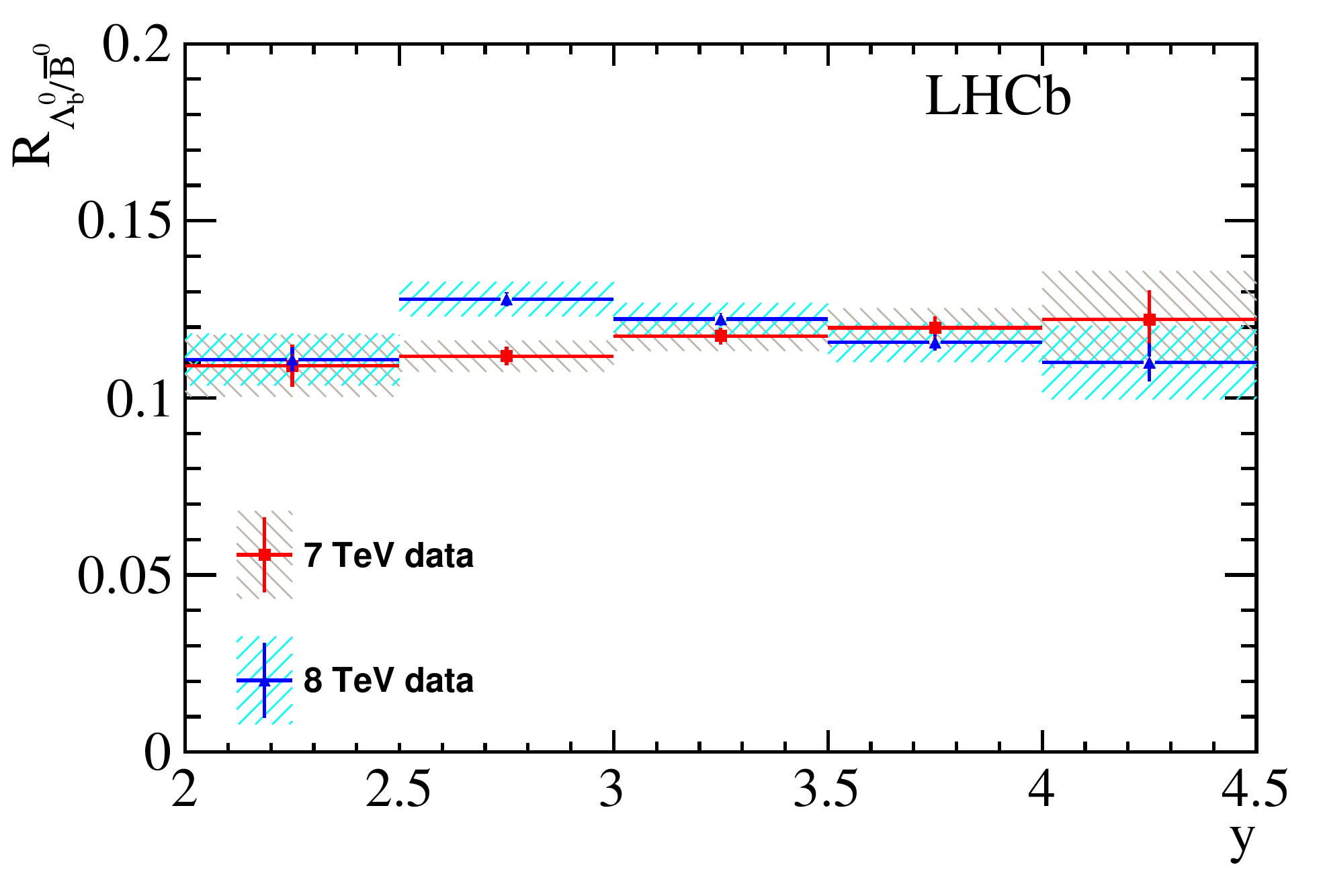}
\caption{ 
    Ratio $\FLbFd$ multiplied by the branching fraction ratio as a function of (left) $\pt$ and (right) $y$ in $pp$
    collisions at  7 and 8 TeV respectively, measured in the $b\to\jpsi X$ decay.
}
\label{fig:figure5}
\end{figure}

The $\Lb$ over $B^0$ production ratio $\FLbFd$ in $pp$ collisions at $\sqrt{s}=7$ TeV is measured with decays
$\Lb\to\Lcp\pi^-, \Lcp\to p K^-\pi^+$ and $\overline{B}^0\to D^+\pi^-, D^+\to K^-\pi^+\pi^+$~\cite{cite3}. 
Experimentally $\FLbFd$ is calculated as $R_{\BR}\times
R_Y$, where $R_\BR$ is the overall branching fraction ratio and $R_Y$ is the ratio of efficiency corrected yields.
However the branching fraction for $\Lb\to\Lcp\pi^-$ is poorly
known, so the analysis measured the kinematic dependence of $\FLbFd$ and used external input of $\FLbFd$, which is
measured with semileptonic decays~\cite{cite4}, to determine $\BR(\Lb\to\Lcp\pi^-)$. The event selection strategy is similar
to the analysis for $f_s/f_d$ and fits to the invariant mass distributions 
are used to obtain the signal yields. The ratio of selection efficiencies is
evaluated using simulated events and validated using real data. The efficiency corrected yield ratio scaled to match
that in semileptonic decays is shown in Figure~\ref{fig:figure4}. The scale factor provides a measurement of
$\BR(\Lb\to\Lcp\pi^-)$, with external inputs of branching fractions for other decays, resulting in
    $\BR(\Lb\to\Lcp\pi^-) = (4.30\pm 0.03^{+0.12}_{-0.11}\pm0.26\pm0.21)\times 10^{-3}$.
Here the uncertainties are statistical, systematic, that from semileptonic analysis and that on the knowledge of
$\BR(\overline{B}^0\to D^+\pi^-)$. From Figure~\ref{fig:figure4}, it can be seen that $\FLbFd$ decreases with the
increase of $\pt$ following approximately an exponential distribution, and it increases linearly with $b$-hadron
$\eta$ within LHCb acceptance.

The kinematic dependence of $\FLbFd$ is as a function of $\pt$ and rapidity is also studied using $\Lb\to \jpsi K^-p$ and $\overline{B}^0\to \jpsi K^-\pi^+$
decay modes~\cite{cite5}, as shown in Figure~\ref{fig:figure5}. The shape of the $\pt$-dependence is consistent with the
measurement using open charm decay modes. The absolute scale provides measurement of the $\Lb\to\jpsi K^-p$ branching
fraction calculated to be $\BR(\Lb\to\jpsi K^-p) = (3.17\pm 0.04\pm 0.07\pm0.34^{+0.45}_{-0.28})\times 10^{-4}$,
where the uncertainties are statistical, systematic, that on $\BR(\overline{B}^0\to\jpsi K^-\pi^+)$ and the
knowledge of $\FLbFd$ respectively.

Table~\ref{tab:table1} summarizes LHCb measurements of $b$-hadron fragmentation functions in semileptonic and
hadronic decay modes. Compiling these numbers, about 75\% of $b$ quarks produced at LHCb fragment into $B^+$ or $B^0$, approximately $10\%$ go
into $B_s^0$ and 15\% result in $\Lb$ baryons.
\begin{table}[t]
\begin{center}
\begin{tabular}{l|ccc}  
    Variables   & Decay modes        &  Results                     & Comments\\ 
    \hline
    $f_s/f_d$~\cite{cite4}   & $D\mu\nu X$   & $0.268^{+0.023}_{-0.021}$    & Independent of $D\mu\,\,\pt,y$ \\
    $2f_{\Lb}/(f_u+f_d)$~\cite{cite4}   & $\Lcp/D\mu\nu X$   & $\sim 40\%$              & Dependent on $D\mu\,\,\pt,y$ \\
    $f_s/f_d$~\cite{cite1}   & $D_s^-\pi^+ \,/\, D^-K^+$   & $0.238\pm0.026$    & Slightly dependent on $b$ $\pt$ \\
    $f_{\Lb}/f_d$~\cite{cite3}   & $\Lcp\pi^-\,/\, D^+\pi^-$   & Fixed for $\BR(\Lb\to\Lcp\pi^-)$ & $f_{\Lb}/f_d$ depends on $b$ $\pt,\eta$\\
    $f_{\Lb}/f_d$~\cite{cite5}   & $\jpsi K^- p \,/\,\jpsi K^- \pi^+$   & Fixed for $\BR(\Lb\to\jpsi K^-p)$& $f_{\Lb}/f_d$ depends on $b$ $\pt$\\
    \hline
\end{tabular}
\caption{ LHCb measurements of fragmentation functions in $pp$ collisions.}
\label{tab:table1}
\end{center}
\end{table}

\section{Measurements of $b$-hadron production asymmetries}
Although $b$ and $\overline{b}$ quarks are produced simultaneously, the production rate of a particular $b$-hadron species is not
expected to be strictly identical to that of its charge conjugate. Since the initial colliding proton-proton is matter
rather than anti-matter, the collision favours the production of $B^+,\,B^0,\,B_s^0$ and $\Lb$ instead of their anti-particles, especially at
large rapidity where this process happens closer to the beam remnants. However as $b$ and
$\overline{b}$ quarks are produced predominately in pairs, the asymmetries of $b$-hadrons should compensate among each other to
make overall zero asymmetry. 

For the LHCb analysis~\cite{cite6}, the production asymmetries for $B^+,\,B^0$ and $B_s^0$ are measured using decays 
$B^+\to \jpsi K^+$, $B^0\to\jpsi K^{*0}$ and $B_s^0\to D_s^-\pi^+$ for $pp$ collisions at 7 and 8 TeV. 
Here the production asymmetry is defined as
$A_\mathrm{p}(x)\equiv\frac{\sigma(\overline{x})-\sigma(x)}{\sigma(\overline{x})+\sigma(x)}$ for
$x=B^+,\,B^0,\,B_s^0,\,\overline{\Lambda}_b^0$, where $\sigma(x)$ stands for the production cross-section. 
Neglecting $B_c^+$ and other $b$ baryons, $A_\mathrm{p}(\Lb)$ is determined from $A_\mathrm{p}(\Bp),\,A_\mathrm{p}(\Bd)$ and $A_\mathrm{p}(\Bs)$ as 
\begin{equation}
    A_\mathrm{p}(\Lb)\approx-\left[\frac{f_u}{f_{\Lb}}A_\mathrm{p}(\Bp) + \frac{f_d}{f_{\Lb}}A_\mathrm{p}(\Bd)  + \frac{f_s}{f_{\Lb}}A_\mathrm{p}(\Bs)
    \right].
    \label{eq:1}
\end{equation}
The raw asymmetry, $A_\mathrm{raw}$, for $\Bp$ hadron is obtained by counting the number of $\Bp$ and $B^-$ signal
yields as obtained from the fit to the invariant mass distribution. 
To determine the production asymmetry, 
the CP asymmetry $A_\mathrm{CP}$ and detection asymmetry $A_D$ are subtracted from the raw asymmetry.
The CP asymmetry for $\Bp$ was measured by LHCb to be
$A_\mathrm{CP}(B^+)=(0.09\pm0.27(\mathrm{stat})\pm0.07(\mathrm{syst}))\times10^{-2}$~\cite{cite7}. The kaon detection
asymmetry is measured using $D$-meson decays produced directly in $pp$ collisions following
the method in reference~\cite{cite8}, while kaon particle identification efficiencies are measured using control
samples.
Due to flavor oscillations between $B^0$ and $\overline{B}^0$, and between $B_s^0$ and $\overline{B}_s^0$, a time-dependent analysis
for $\Bd$ and $\Bs$ is required to measure $A_\mathrm{p}(B_{d,s})$. Summing over the two initial flavors, the observed time-dependent production rate reads
\begin{eqnarray}
    S(t,\psi)\propto
    [1-\psi(A_\mathrm{CP})]\{e^{-\Gamma_{d(s)}}[\Lambda_+\cosh(\Delta\Gamma_{d(s)}t/2)+\psi\Lambda_-\cos(\Delta
    m_{d(s)}t)]\otimes R(t)\}\epsilon(t), \nonumber\\
    \mathrm{with}\,\, \Lambda_{\pm}\equiv(1-A_\mathrm{p})|q/p|^{1-\psi}\pm(1+A_\mathrm{p})|q/p|^{-1-\psi}\nonumber,
\end{eqnarray}
where $R(t)$ and $\epsilon(t)$ describe detection resolution and efficiency, $\Gamma$, $\Delta\Gamma$, $\Delta m$ and $|q/p|$ are the
oscillation parameters taken as external inputs, and $\psi=\pm1$ tags the $B_{d,s}$ flavor at decay using the final states. From
the above equation, it can be seen that the production asymmetry is measured from the asymmetric patterns
of the decay time distributions between $B_{d,s}$ and $\overline{B}_{d,s}$, and
$A_\mathrm{CP}$ and $A_D$ factor out in the measurement. 

\begin{figure}[!htbp]
\centering
\includegraphics[width=0.24\textwidth]{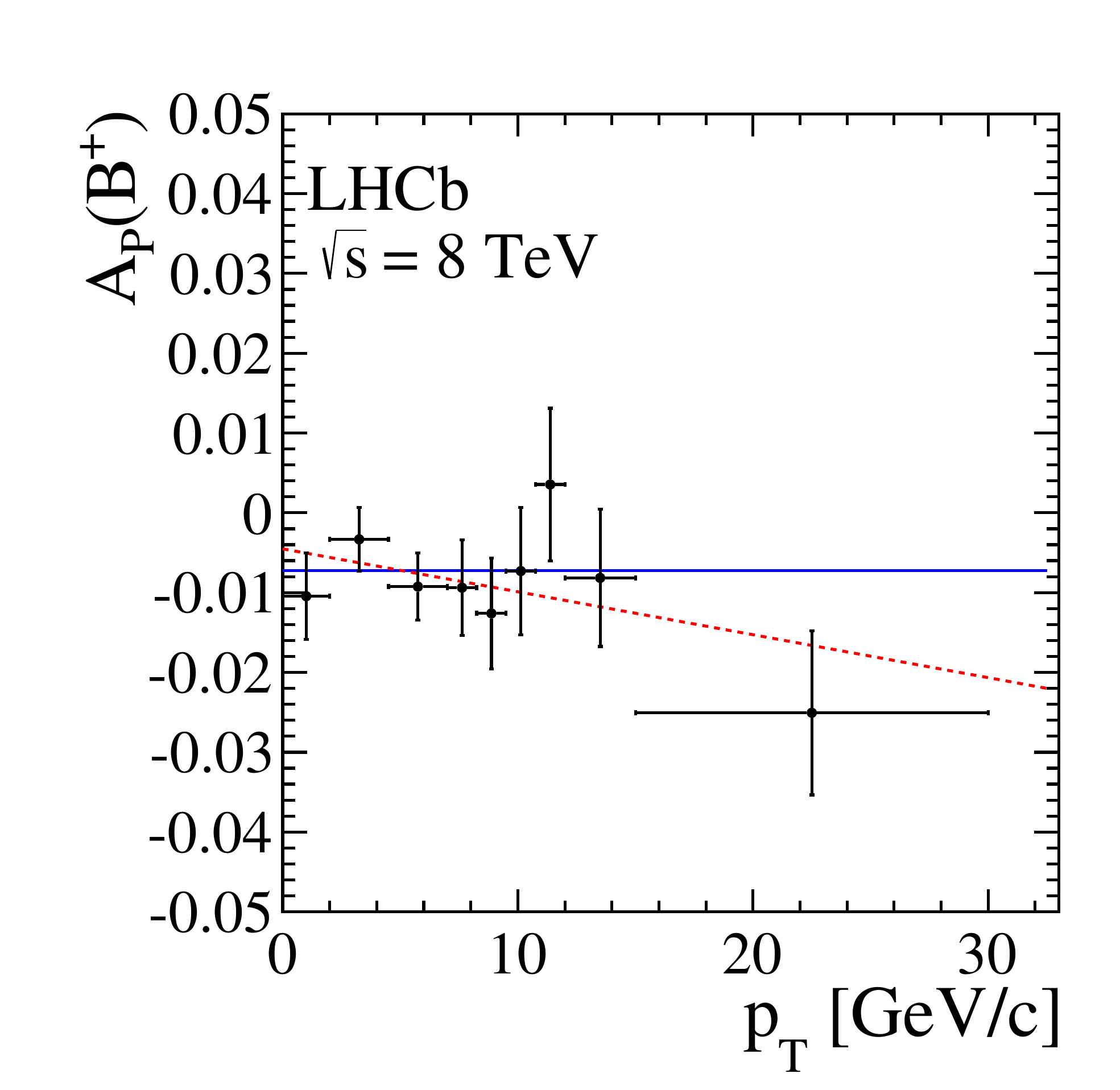}
\includegraphics[width=0.24\textwidth]{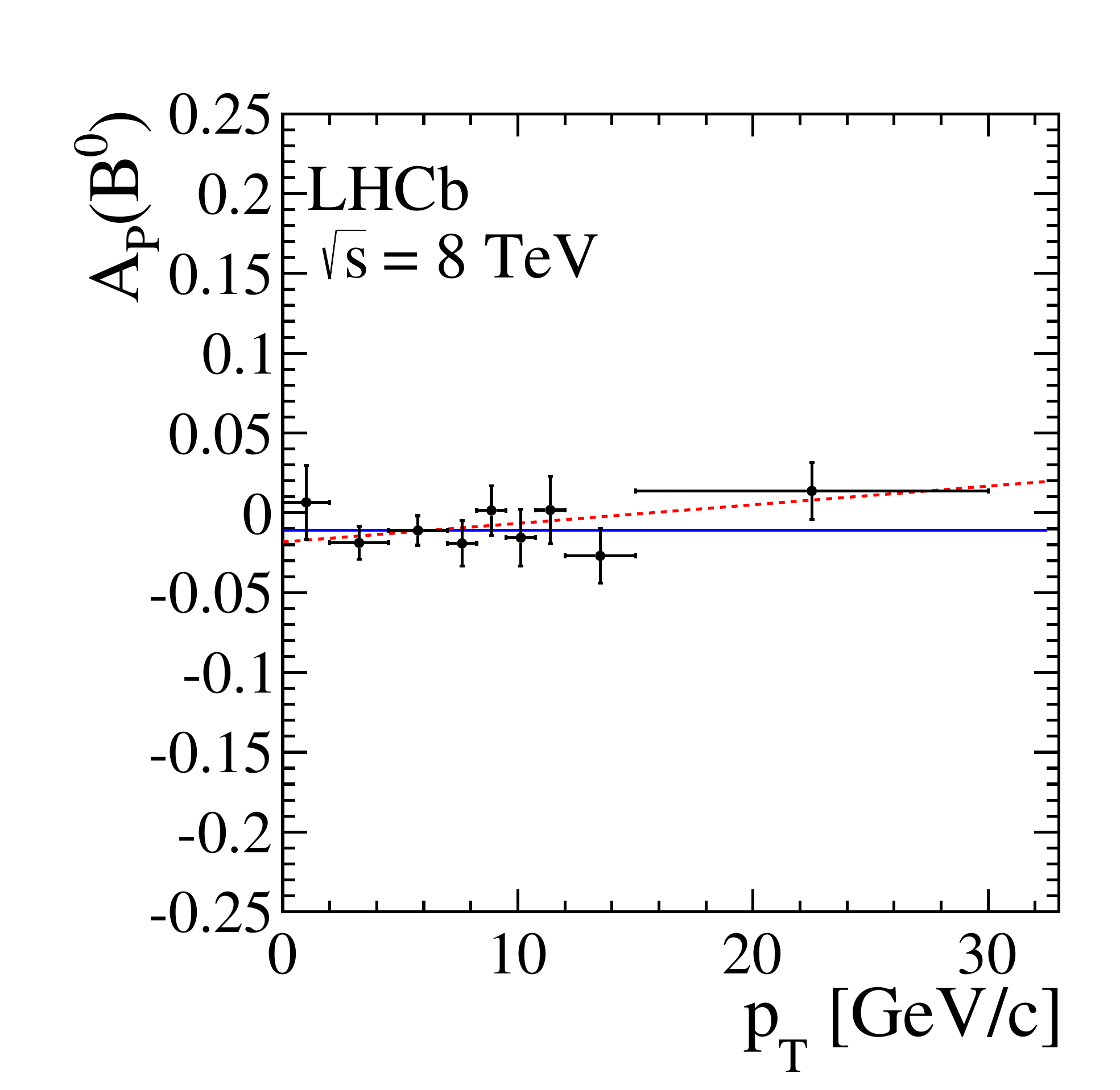}
\includegraphics[width=0.24\textwidth]{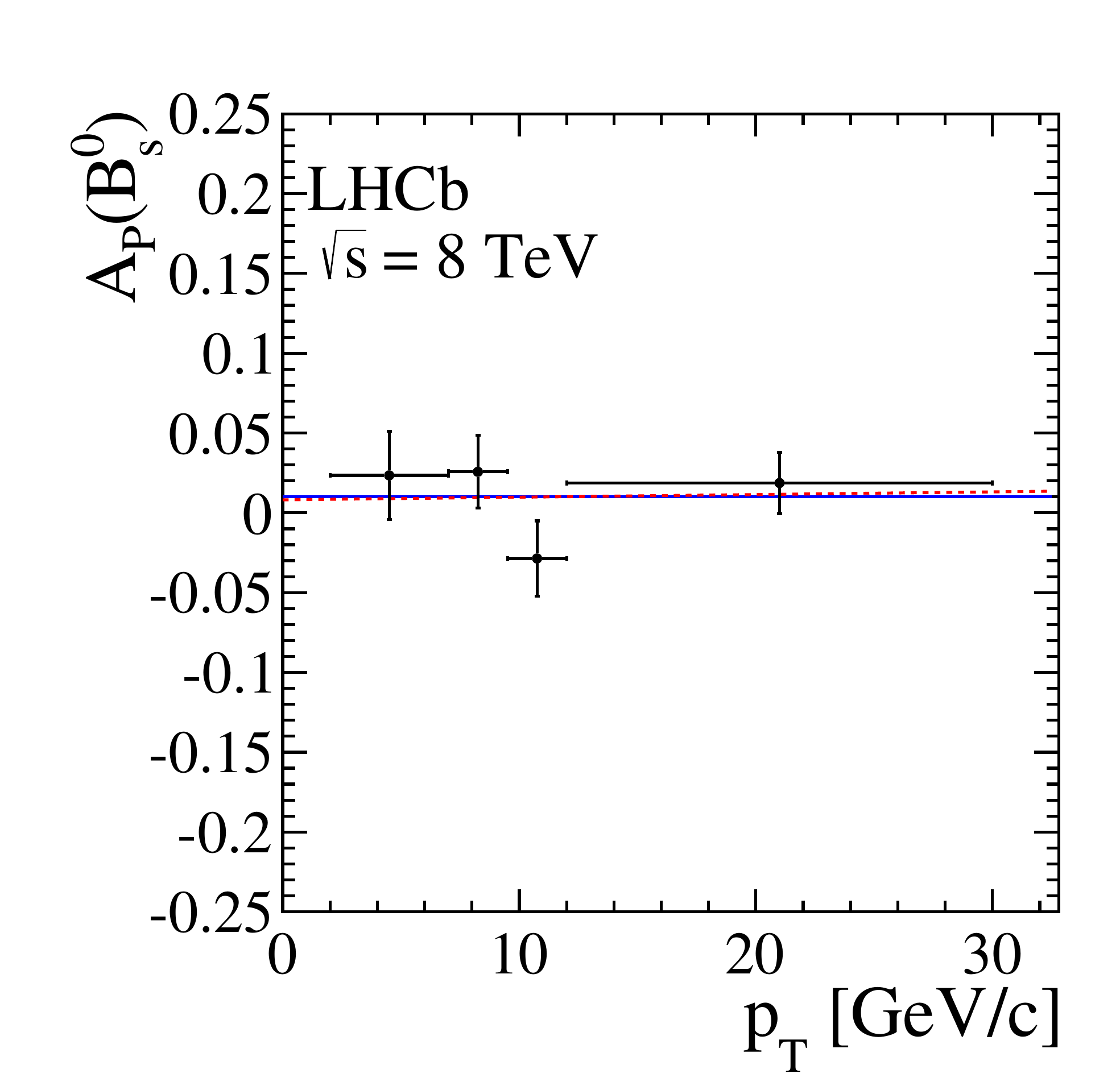}
\includegraphics[width=0.24\textwidth]{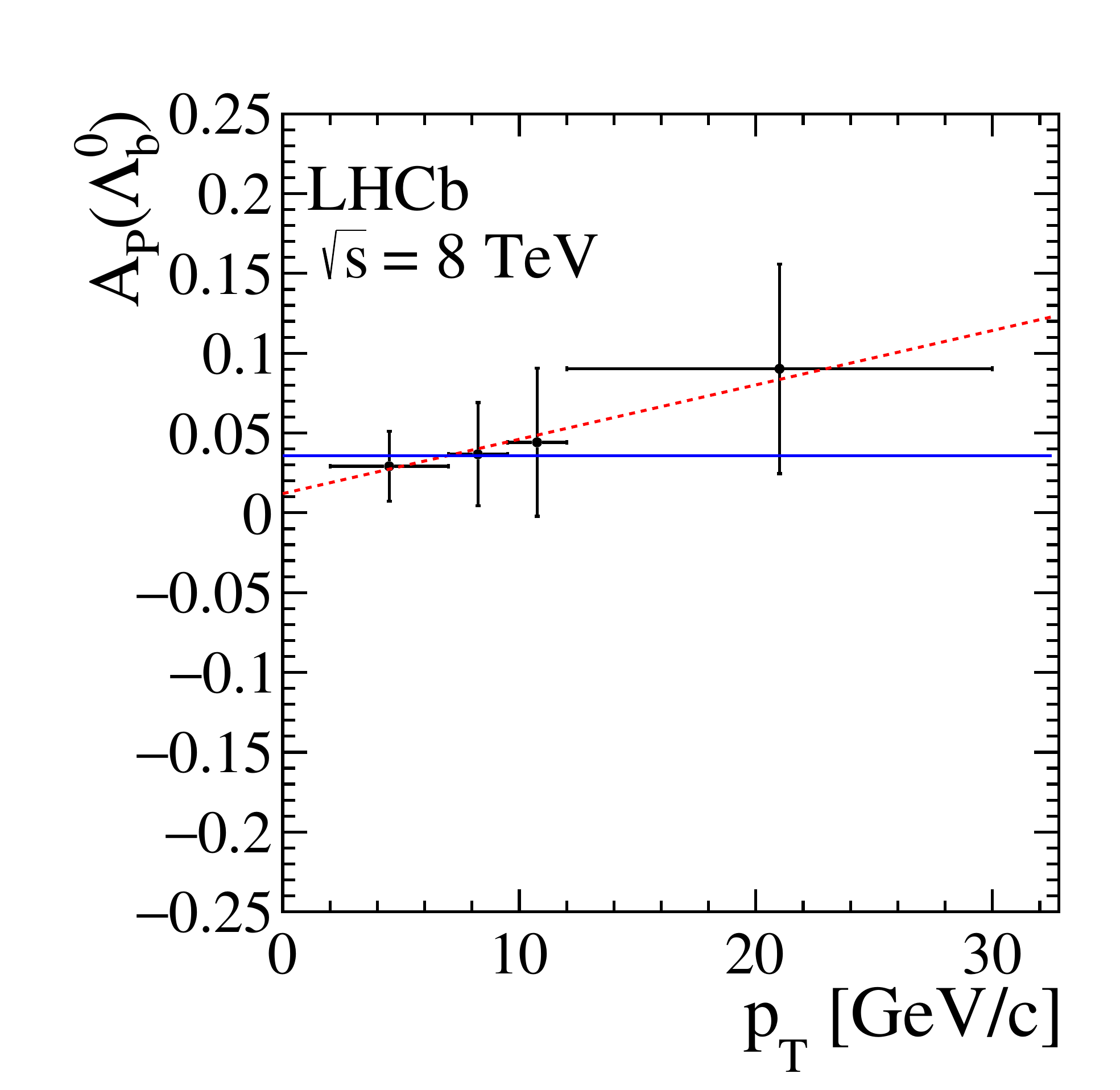}
\includegraphics[width=0.24\textwidth]{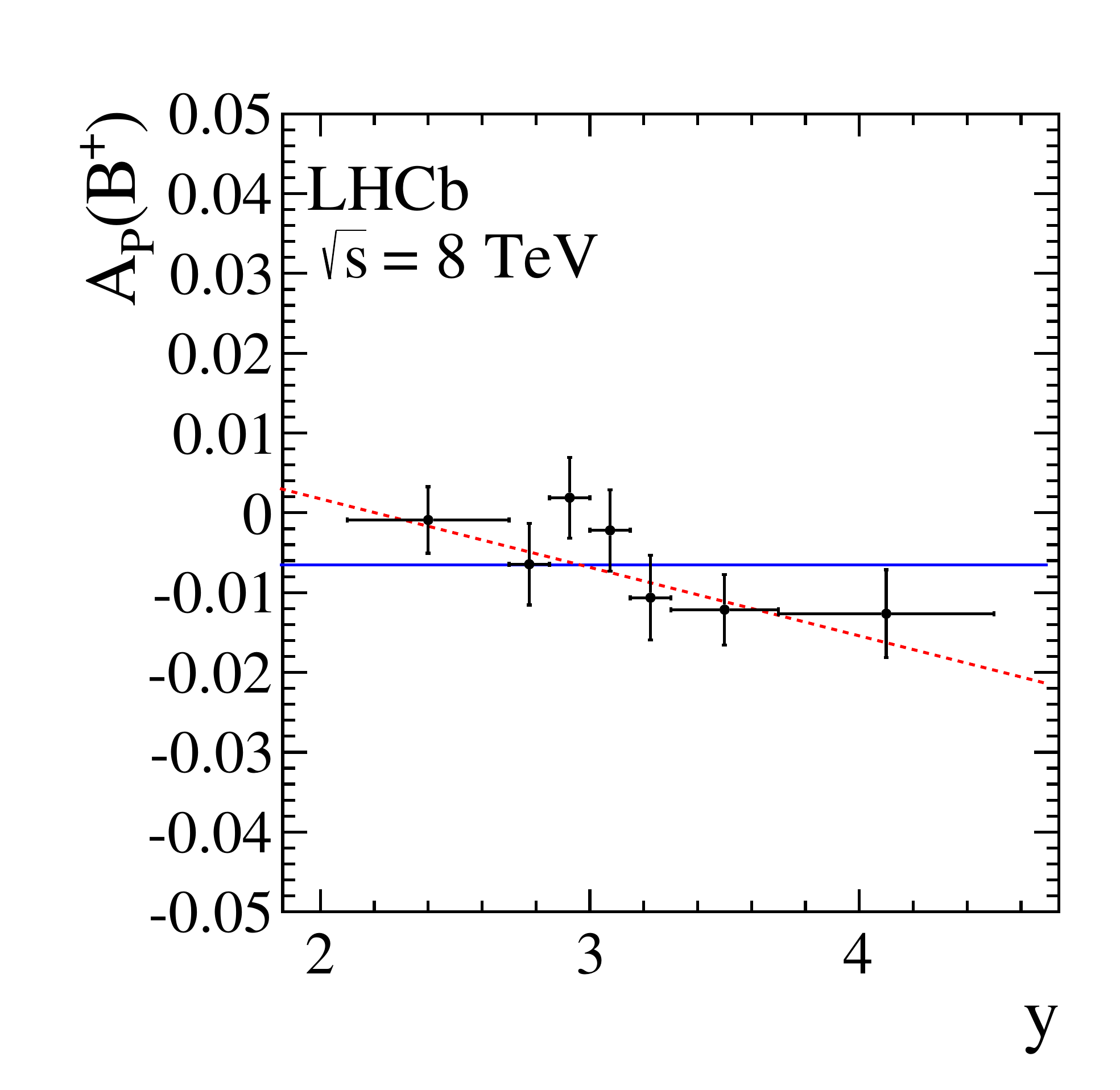}
\includegraphics[width=0.24\textwidth]{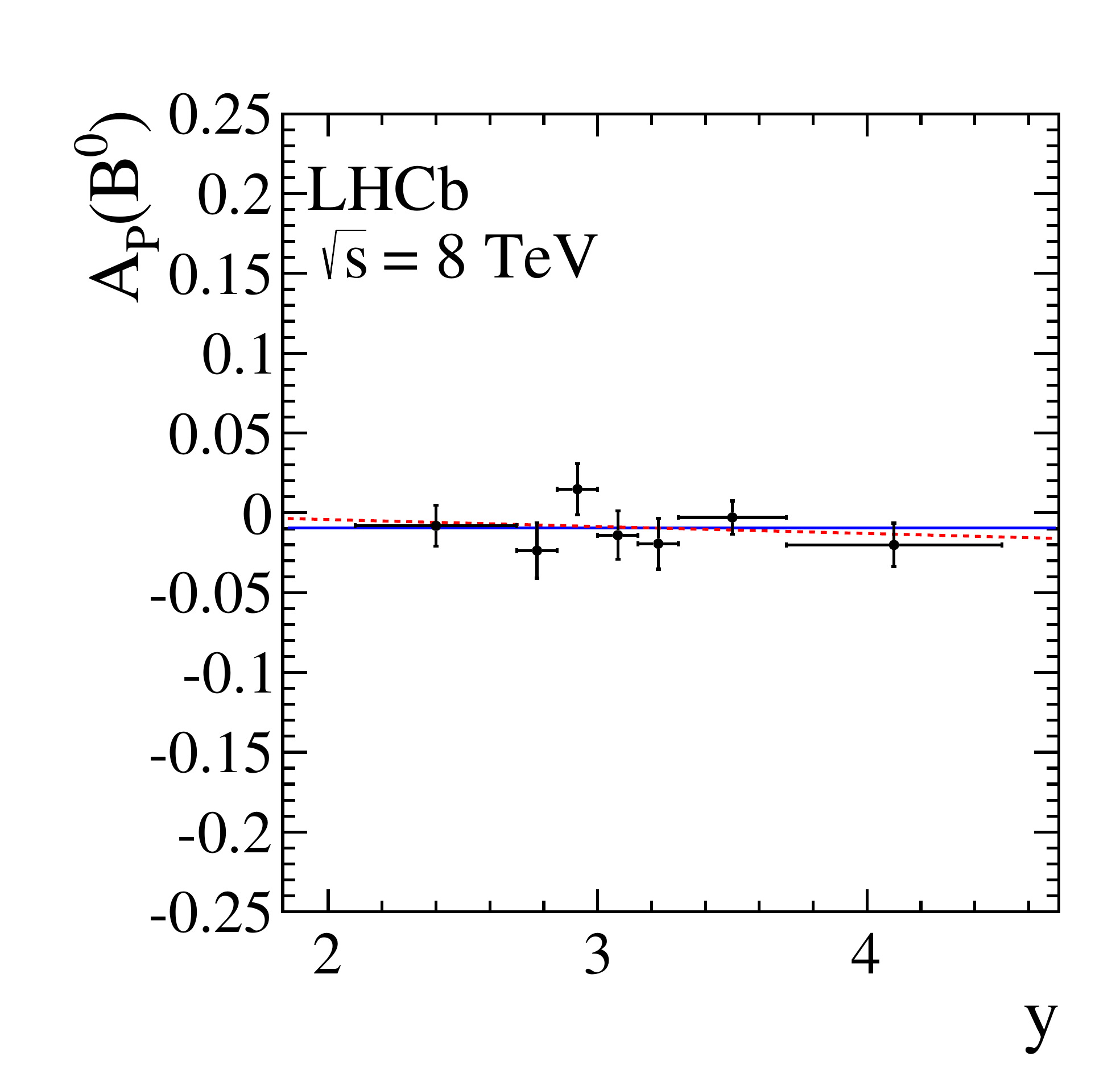}
\includegraphics[width=0.24\textwidth]{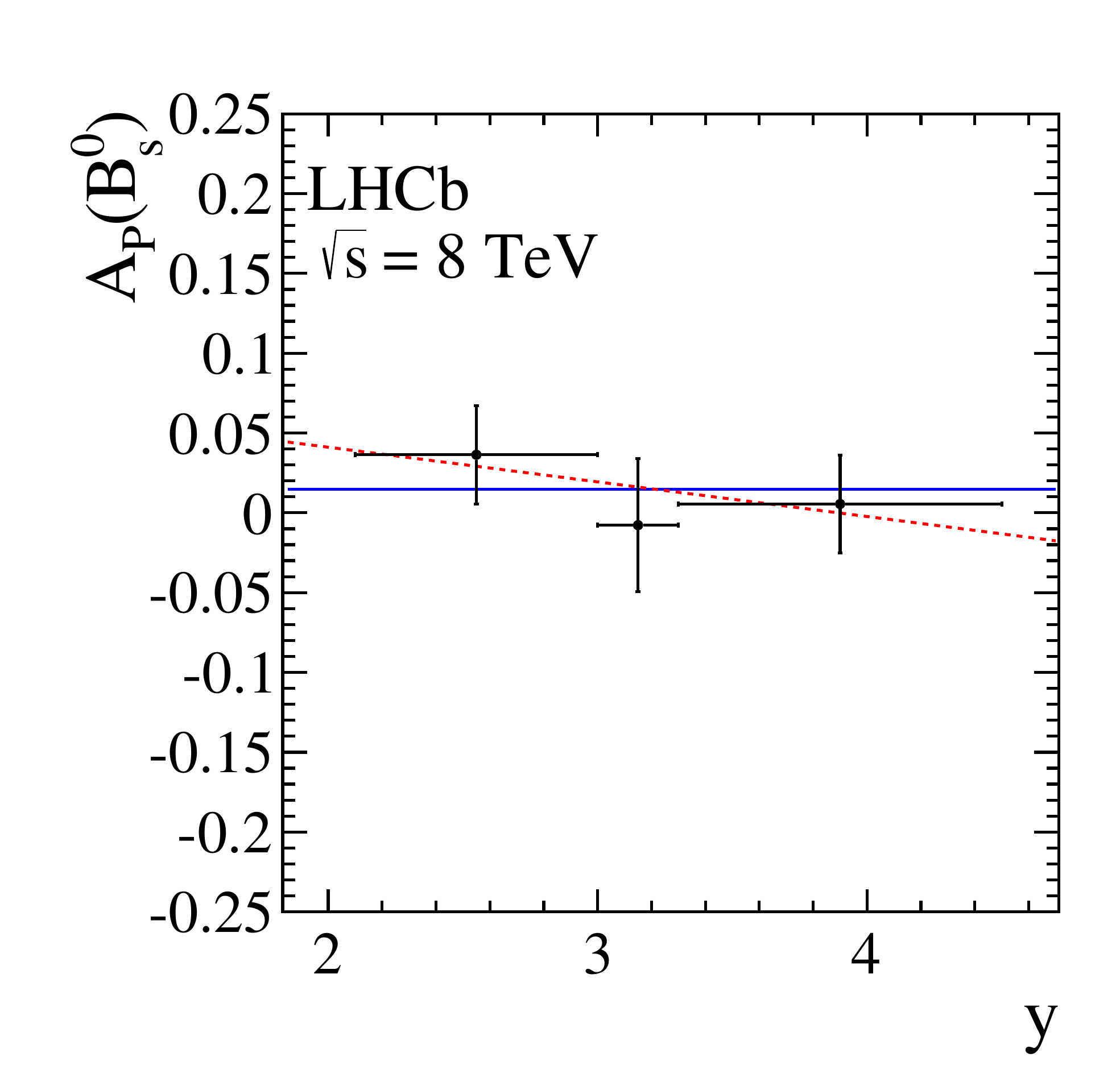}
\includegraphics[width=0.24\textwidth]{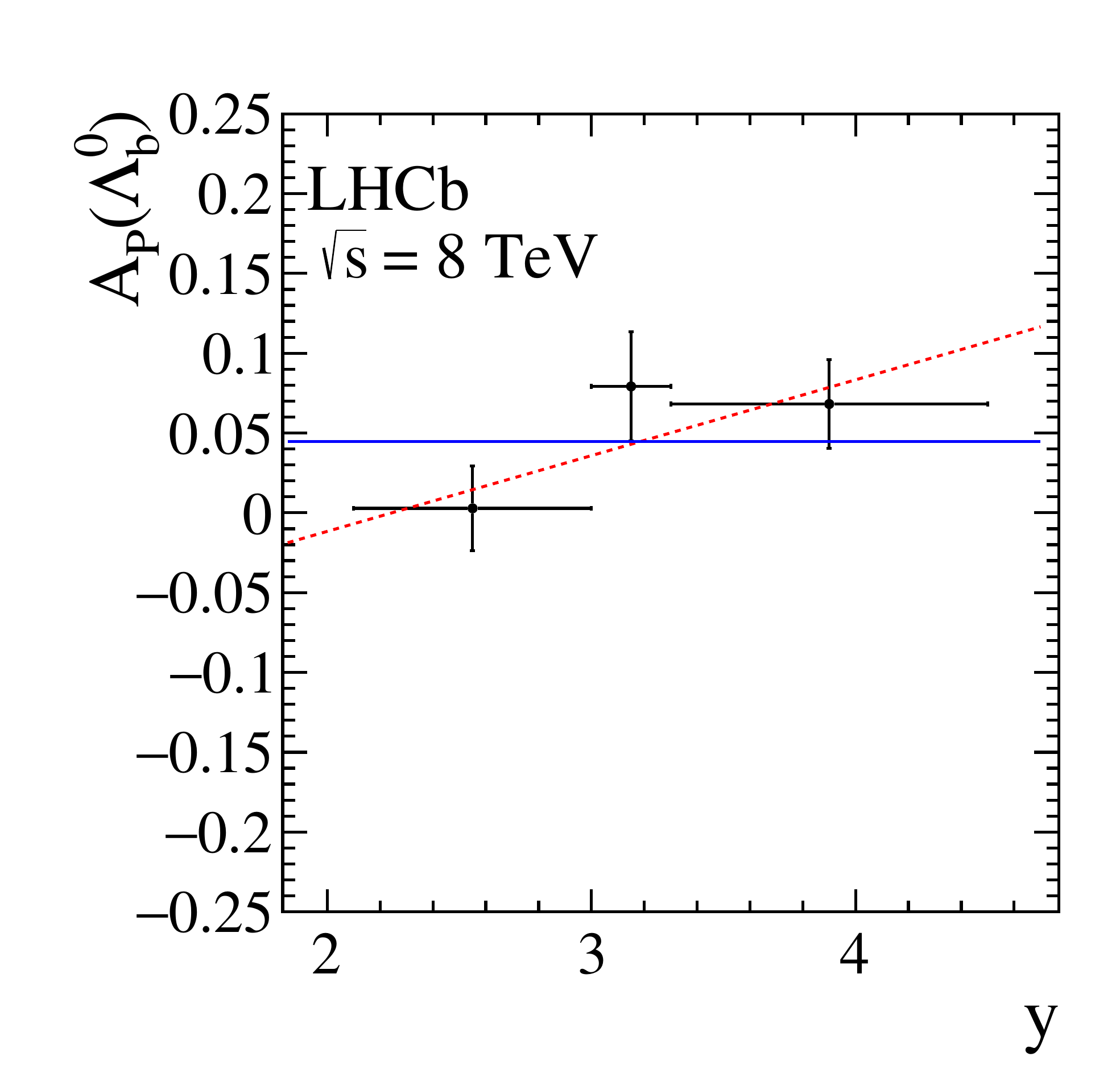}
\caption{ 
    The measured production asymmetries of $\Bp$, $\Bd$, $\Bs$ and $\Lb$ as functions of (top row) $b$-hadron $\pt$ and (bottom
    row) $y$ in $pp$ collisions at 8 TeV.
}
\label{fig:figure7}
\end{figure}

The measurements of production asymmetries for $\Bp$, $\Bd$ and $\Bs$ are then used obtain $A_\mathrm{p}(\Lb)$ using
Equation~\ref{eq:1}.
For the results on $A_\mathrm{p}$, the systematic uncertainties for $A_\mathrm{p}(\Bp)$ and $A_\mathrm{p}(\Bd)$ are
dominated by external inputs of
$A_\mathrm{CP}(B^+)$ and the oscillation parameter respectively, and that for $A_\mathrm{p}(\Bs)$ is dominated by 
invariant mass and decay time fits. The uncertainties on $A_\mathrm{p}(\Bp)$, $A_\mathrm{p}(\Bd)$ and $A_\mathrm{p}(\Bs)$ propagate to $A_\mathrm{p}(\Lb)$
measurement in addition to the uncertainty of $\Lb$ fragmentation function and uncertainty due to neglecting other $b$-hadrons.
The results suggest that production asymmetries for all the four $b$-hadrons are consistent with zero within 2.5
standard deviations.
The $A_\mathrm{p}$ distributions as functions of $\pt$ and $y$ are displayed in Figure~\ref{fig:figure7} for the 8 TeV dataset, showing kinematic
dependences consistent with zero within uncertainties. There is a hint that $A_\mathrm{p}(\Lb)$ slightly depends on
the rapidity.

\begin{figure}[!htbp]
\centering
\includegraphics[width=0.4\textwidth]{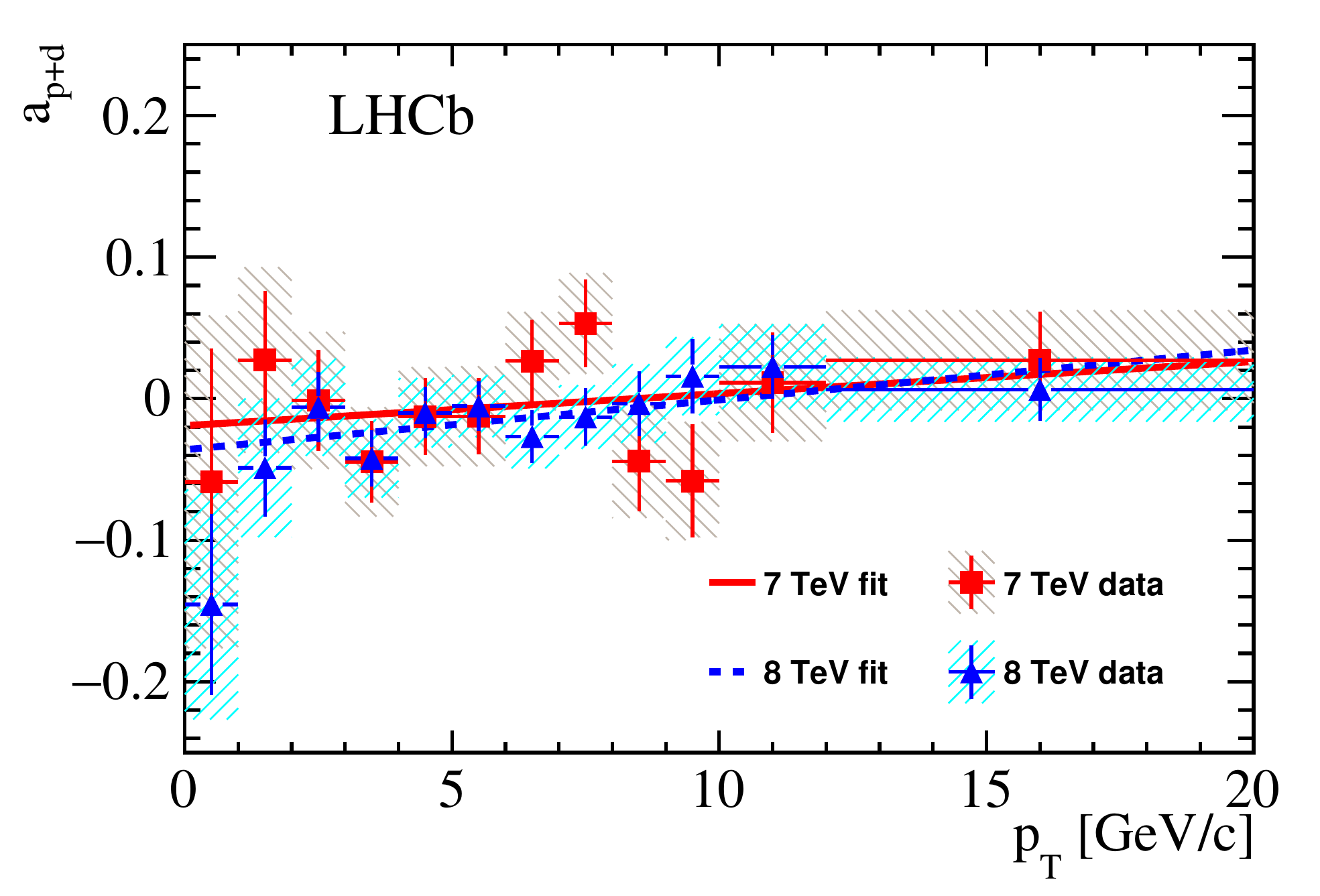}
\includegraphics[width=0.4\textwidth]{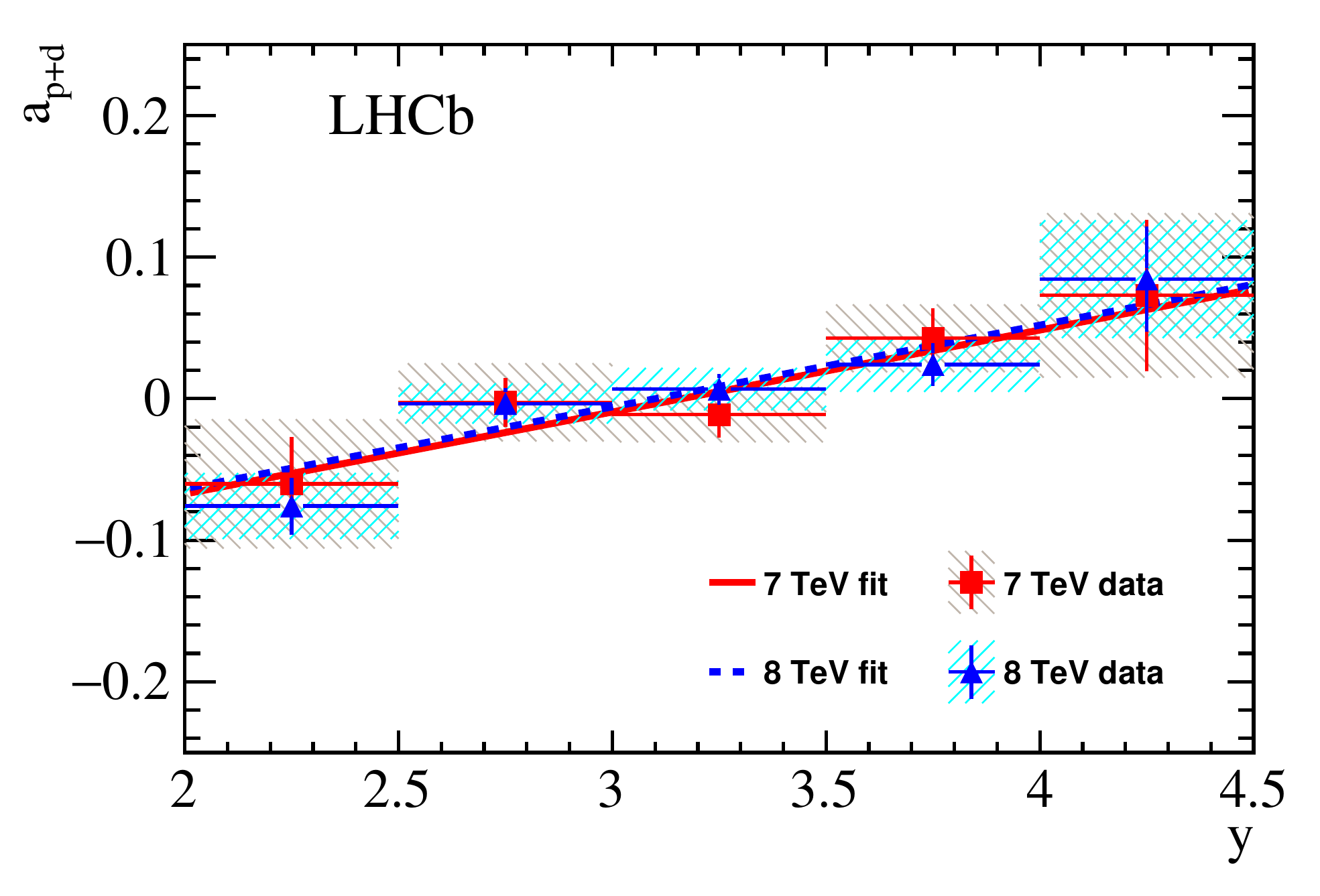}
\caption{ 
    The production plus CP asymmetry of $\Lb$ measured using $\Lb\to\jpsi K^-p$ decay as a function of (left) $\pt$ and
    (right) $y$ respectively.
}
\label{fig:figure8}
\end{figure}

The $\Lb$ production asymmetry is also measured directly using the $\Lb\to\jpsi K^- p$ decay mode~\cite{cite5}. For the
measurement the CP asymmetry $A_\mathrm{d}$ in the decay is not disentangled due to the lack of external input. The result is shown in Figure~\ref{fig:figure8} as a function of $\pt$
and $y$ respectively. While the overall asymmetry is consistent with zero, there is an strong evidence of dependence on $\Lb$
rapidity, which is measured to be $A_{p+\mathrm{d}} = (-0.001\pm0.007) + (0.058\pm0.014)\times(y-3.1)$ by fitting the
$A_{p+\mathrm{d}}$ distribution with a linear function.

\section{Conclusions}
The LHCb experiment has made significant contributions to heavy flavor production measurements, including the
fragmentation functions for $\Bp$, $\Bd$, $\Bs$ and $\Lb$ hadrons, and their production asymmetries in $pp$
collisions. The results provide inputs for measurements of absolute branching fractions and CP violation in $b$-hadron
decays respectively. With LHCb run 2 data, many new measurements concerning heavy flavor productions at 13 TeV are expected.

\Acknowledgements
The corresponding author acknowledges support from the European Research Council (ERC)
through the project EXPLORINGMATTER, funded by the ERC through a ERC-Consolidator-Grant.

\end{document}